\let\pdfoutput\defined

\documentclass{PoS}
\usepackage{amsmath, amsfonts, bbm, dsfont, mathrsfs,amssymb}
\usepackage{dcolumn}

\title{%
\vspace*{-1cm}
\begin{minipage}{\textwidth}
\begin{flushright}
\texttt{\footnotesize
PoS(LAT2009)005\\%
FERMILAB-CONF-09-554-T\\%
}
\end{flushright}
\end{minipage}\\[15pt]
Light Hadron Masses and Decay Constants%
}

\ShortTitle{Light Hadron Masses and Decay Constants}

\author{\speaker{Enno E. Scholz}%
        \\
        Fermi National Accelerator Laboratory%
        \thanks{Operated by Fermi Research Alliance, LLC, under Contract No.~DE-AC02-07CH11359 with the United States Department of Energy.}%
        , Batavia, IL 60510, USA\\
        E-mail: \email{scholzee@fnal.gov}\thanks{after October 1, 2009: Institut f\"ur Theoretische Physik, Universit\"at Regensburg, 93040 Regensburg, Germany, E-mail: enno.scholz@physik.uni-regensburg.de}}

\abstract{%
The extraction of the light hadron spectrum from a first-principle Quantum Chromodynamics approach is a profound application for lattice simulations of Quantum Chromodynamics.

This review will cover recent lattice results for the masses and decay constants of the light hadrons. In particular, the applicability of different approaches for the extrapolation towards the physical point will be discussed.%
}

\FullConference{The XXVII International Symposium on Lattice Field Theory\\
		 July 26-31, 2009\\
		 Peking University, Beijing, China}

\def\nicefrac#1#2{\leavevmode\kern.1em\raise.5ex\hbox{\the\scriptfont0 #1}\kern-.1em/\kern-.15em\lower.25ex\hbox{\the\scriptfont0 #2}}

\newcolumntype{C}{>{$}c<{$}}
\newcolumntype{R}{>{$}r<{$}}
\newcolumntype{L}{>{$}l<{$}}

\newlength{\closercaption}
\newlength{\afterTable}
\newlength{\afterFigure}
\newlength{\closersection} 
\begin{document}


\noindent
As of today it is widely believed that the theory of the strong interactions in the Standard Model is given by Quantum Chromodynamics (QCD). Hadrons are formed as bound states of different quark flavors, with the gluons mediating the strong interactions. Due to the strong coupling of QCD, a perturbative approach is not feasible at low energies, where the light hadron spectrum is observed in Nature. Simulations of QCD performed numerically on a finite lattice discretizing space and time offer the possibility to study the strong interactions non-perturbatively from a first-principle approach. In this review, I will try to summarize recent achievements and efforts made in lattice QCD simulations to extract the masses and decay constants of the light hadrons, i.e., hadrons built from the three lightest quark flavors---the up, down, and strange quarks. For a review on heavy quark physics on the lattice see the review by Christopher Aubin \cite{Aubin:2009yh} given at the Lattice 2009 conference in Beijing. See also the reviews by Yasumichi Aoki \cite{Aoki:2010yq}, Chulwoo Jung \cite{Jung:2010jt}, Vittorio Lubicz \cite{Lubicz:2010nx}, and Ruth Van de Water \cite{VandeWater:2009uc} , which naturally have some overlap with the topics covered here and for the status as of last year's conference, see the review given by Karl Jansen at that time \cite{Jansen:2008vs}.

This review is structured as follows: first I will discuss the light pseudo-scalar meson sector, paying special attention to chiral perturbation theory, which is used to guide the extrapolations. Besides discussing the extrapolations for the pseudo-scalar meson masses and decay constants, I will also cover results obtained for the light quark masses and for some of the low-energy constants of chiral perturbation theory. In Section~\ref{sec:hadronSpect} I will broaden the discussion to cover the complete light hadron spectrum. Again, now with emphasis on the baryon masses, the techniques for the extrapolation towards the physical point will be reviewed. The status of studies focussing on excited states in the light hadron spectrum and related topics are summarized in the remainder before I give some concluding remarks.


\section{\label{sec:meson} Masses and decay constants of the light pseudo-scalar mesons}

To extract the masses and decay constants of the light pseudo-scalar mesons---the neutral and charged pions and kaons---(for the light vector-meson ($\rho$) see Sec.~\ref{subsec:rho}) from most current simulations with dynamical quarks, an extrapolation to the physical pion mass point still has to be performed. Nowadays a typical value for the lightest dynamical meson mass is 250 MeV, but first results exist for dynamical meson masses in the range of 160 to 190 MeV. The inclusion of dynamical fermions has become standard due to improved algorithms developed over the last few years and increased computing power available, leaving the quenched approximation, with its undetermined systematic error, obsolete for almost all investigations of lattice QCD. In most lattice QCD simulations either two or three dynamical quark flavors are included, although first studies with four dynamical fermions are being pursued as well. The two lightest quarks, the up and the down quark, are usually taken as mass-degenerate, while the third quark flavor, the strange quark, is assigned a heavier mass. For that reason it is common to speak of either $N_f=2$ or $N_f=2+1$ dynamical quark flavor simulations. Here I will focus mainly on aspects related with the extraction of quantities at the physical values of the quark masses. In the parameter space of quark masses, this point is usually defined by the constraint that the masses of the pion and kaon, $m_\pi$ and $m_K$, (the latter only if the strange quark is considered, i.e.\ in simulations with $N_f=2+1$ flavors) take their experimentally observed masses at that point. To define this point, besides $m_\pi$ and $m_K$ a third quantity is needed to set the overall scale in the computation. Here some groups utilize the mass of the $\Omega^-$ baryon, because of its expected weak dependence on the up and down quark masses. Other choices include the pion decay constant and the scale parameters $r_0$ (Sommer scale) or $r_1$ extracted from the static quark potential. These different choices also provide opportunities for cross-checks, which are needed since none of the above choices is believed to be unproblematic in current simulations. For example, the Sommer scale $r_0$ is not known a priori from experiment, and different estimates from lattice simulations vary by as much as 10 per cent. In the case of the pion decay constant, the extrapolation to the physical point as discussed below may introduce some unwanted systematic uncertainty. Recently, the $\Omega^-$ baryon mass has been advertised to overcome these difficulties, since it is a experimentally well determined quantity with only mild dependence on the up and down quark masses, being made from three strange quarks. But it remains to be seen, how big finite volume effects are in the case of this heaviest baryon within the light baryon spectrum. 

Table~\ref{tab:chPTfits} gives an overview of recent dynamical simulations, a detailed discussion will follow in Sec.~\ref{subsec:chPTanalyses}. As one can see, several fermion formulations are used, which differ in their approach to the continuum limit and chiral properties. Wilson fermions and their improved versions are cheap to simulate but introduce additional chiral symmetry breaking due to lattice artifacts, twisted mass fermions can be seen as a special variant of improved Wilson fermions. The (improved) staggered fermion formulations are very cheap to simulate and leave a remnant of the chiral symmetry unbroken at the expense of introducing additional taste degrees of freedom, which have to be accounted for. A more improved chiral behavior at the expense of additional simulation cost is offered by the domain wall formulation for fermions or the overlap formulation. The latter even offers exact chiral symmetry at finite lattice spacing but is also most computationally demanding. I will not go into further technical details of the ensemble generation or a cost comparison between these simulations. See the review of current dynamical simulations given by C.~Jung at this conference \cite{Jung:2010jt} and references therein. All except one analysis is currently relying on chiral perturbation theory to perform the extrapolation from their simulated meson masses to the physical pion and kaon mass: the PACS-CS collaboration in their recent work followed a different approach, namely to reweight their ensemble generated with a lightest meson mass of approximately 160 MeV to the physical pion mass \cite{Kuramashi:LAT09}.

%
%
\begin{table}
\begin{center}
\begin{tabular}{*{7}{c}}
\hline\hline
   & $N_f$ & $a$ [fm] & $m_{\rm PS}$ [MeV] & ${\rm N}^n{\rm LO}$ & & $m_{\rm PS}L$ \\\hline
ETMC  & 2 twisted m. & 0.08, 0.07,              & 250--600 & 2 & compl.\ SU(2) & $\geq3.0$ \\
      &              & 0.05 $\to$ 0\\[10pt]
JLQCD & 2 overlap    & 0.12                     & 290--750 & 2 & ``$\xi$'' SU(2) & $\geq 2.8$ \\
      & 2+1 overlap  & 0.10                     & 320--800 & 2 & ``$\xi$'' SU(2),SU(3) & $\geq 2.8$ \\[10pt]
PACS-CS & 2+1 iWilson & 0.09                    & 160--410 & 1 & compl.\ SU(2) & $\geq2.3$ \\
        &             &                         & \multicolumn{4}{l}{reweight $\to m_\pi$} \\[10pt]
MILC    & 2+1 staggered & 0.09,0.06,            & 180--380 & 1 & rS$\chi$PT SU(3) & $\geq 4.0$  \\
        &               & 0.045 $\to$ 0         &          & 2 & compl.\ SU(3) \\
        &               &                       & 180--540 & 3,4 & analytic \\[10pt]
Aubin et al. & 2+1 stagg/DWF & 0.12,0.09        & 240--500 & 1 & SU(3) MA$\chi$PT& $\geq 4.0$ \\
        &               &   $\to$ 0                    &          & $\geq 3$ & analytic & \\[10pt]
RBC-UKQCD & 2+1 DWF & 0.11, 0.09                & 290--420 & 1,2 & compl.\ SU(2) &  $\geq 4$ \\
          &         & $\to$ 0 \\\hline\hline
\end{tabular}
\end{center}
%
\newsavebox{\refer}
\sbox{\refer}{\small Sec.~\ref{subsec:chPTanalyses}}
\caption{\label{tab:chPTfits}Overview of chiral fits for the light meson masses and decay constants performed by various groups. The table lists the number of dynamical fermion flavors $N_f$ included in the simulation and the fermion discretization used (two in the case of a mixed approach) and the lattice spacing(s). An entry $a \to 0$ indicates that the continuum limit has been taken at some point in the analysis. $m_{\rm PS}$ specifies the range of the dynamical lightest meson masses (lower partially quenched meson masses may have been included in the analysis as well). The specific $\chi$PT extrapolation applied and the order thereof is indicated by the exponent $n$ in N$^n$LO. To indicate the importance of finite volume corrections (which have been included in all cases), the lower bound for $m_{\rm PS}L$ is provided, too. See \usebox{\refer} for references and details.}
\end{table}


\subsection{\label{subsec:chPTmeson}Chiral perturbation theory for the meson sector}

Chiral perturbation theory ($\chi$PT) is an effective theory to describe the spontaneous and explicit breaking of chiral symmetry as an expansion in the masses and momenta of the light mesons \cite{Weinberg:1978kz,Gasser:1983yg,Gasser:1984gg}, see, e.g., \cite{Bijnens:2006zp,Bijnens:2009mw} for recent reviews. The mesons act as the fundamental fields in this theory. Depending on whether one assumes chiral symmetry in the massless limit of two (up, down) or three (up, down, strange) quark flavors (usually referred to as the chiral limit), one formulates either ${\rm SU}(2)\times{\rm SU(2)}$ or ${\rm SU}(3)\times{\rm SU}(3)$ $\chi$PT (in the following I will for short just write ${\rm SU}(2)$ or ${\rm SU}(3)$ $\chi$PT). In addition to spontaneous breaking and explicit breaking (due to the non-zero quark masses) of the chiral symmetry from ${\rm SU}(N)_L\times{\rm SU}(N)_R$ to ${\rm SU}(N)_V$, also lattice artifacts can introduce additional symmetry breaking effects. It is possible to address the latter in the $\chi$PT analyses. Therefore a distinction between continuum and lattice $\chi$PT for a specific formulation of the fields on the lattice has to be made. Besides the quark mass parameters%
\footnote{I will only deal with the case of two mass-degenerate light quarks, since this is current practice in lattice simulations, although $\chi$PT formulae for quarks with non-degenerate masses are available in the cited literature as well.}%
 $m_{\rm ud}=(m_{\rm u}+m_{\rm d})/2$ and $m_{\rm s}$, the effective Lagrangian of $\chi$PT also depends on several low-energy constants (LECs, sometimes referred to as Gasser-Leutwyler coefficients) which include effects of the heavier quark flavors as well as high energy QCD interactions. In lowest order two LECs appear: $B$ and $f$ for SU(2) or $B_0$ and $f_0$ for SU(3). The latter being the decay constant in the chiral limit (here I choose a normalization, such that the physical $f_\pi\approx 130\,{\rm MeV}$). In higher orders, additional constants appear, usually denoted as $L_i$ and $K_i$. One has to keep in mind, that the LECs of SU(2) and SU(3) differ, since the former include the effects of the strange quark as well.

Lattice simulations, besides relying on $\chi$PT for the extrapolation to the physical point, are also able to provide valuable information on $\chi$PT. In contrast to experimental measurements where the quark masses are necessarily fixed to their values in Nature, they can be freely varied in lattice QCD simulations. Therefore, lattice QCD should be able to test the predictions of $\chi$PT as functions of the quark masses and---if all lattice systematics are well understood and under control---make predictions about the convergence region and extract low-energy constants relevant for phenomenological models.

A natural choice would be to use SU(3) $\chi$PT for  2+1 dynamical flavor simulations and SU(2) for those with only 2 flavors. However, it turned out that the convergence at next-to-leading order (NLO) in SU(2) $\chi$PT is much better than that in SU(3) at NLO. Especially, it is questionable, whether the meson masses close to or above the physical kaon mass can be successfully described by NLO SU(3) $\chi$PT. This has first been observed by the RBC-UKQCD Collaboration \cite{Allton:2008pn} and was later confirmed by other groups, e.g.\ \cite{Aoki:2008sm}. When applying SU(2) $\chi$PT to data from $N_f=2+1$ simulations, the LECs are obtained at the simulated heavy quark mass. Practically, this is not a drawback, since the simulated heavier quark mass can be tuned to lie close enough to the physical strange quark mass, usually within 10 to 15 per cent is achieved. To account for the remaining mismatch in the heavy quark mass tuning, either simulations at different heavy quark masses and a subsequent interpolation have to be performed, or reweighting in the heavy quark mass might offer a promising remedy \cite{Kelly:2009fp}. 

Table~\ref{tab:chPTfits} also lists the different variants of $\chi$PT, which have been used in the analyses of the various groups. Besides the continuum $\chi$PT there are variants which include certain effects due to the fermion discretization on the lattice like rooted Staggered $\chi$PT (rS$\chi$PT) and Mixed Action $\chi$PT (MA$\chi$PT). The former takes into account the rooting procedure necessary in the case of staggered fermions, while the latter is applicable when different fermion actions are used in the sea and valence sector. I did not explicitly mention the use of partially quenched $\chi$PT (PQ$\chi$PT) in the table. Partially quenching refers to the situation, when in addition propagators with quark masses different from the dynamical masses are calculated. Those partially quenched quarks are referred to as the valence sector in contrast to the sea sector of the dynamical (i.e. unquenched) quark masses.

In addition to these $\chi$PT based extrapolations, there are also several ones which could be best described as $\chi$PT-inspired. Currently, the JLQCD Collaboration uses fit formulae based on a resummation argument, where in NLO and beyond the LO squared meson mass $2B m_q$ gets replaced by the measured meson mass squared ($m_{\rm PS}^2$) and likewise the decay constant in the chiral limit $f$ by the measured one $f_{\rm PS}$. Since they also replace the scale $\mu$ in the chiral logarithms introduced by the regularization%
\footnote{For example, a typical term beyond LO of the form $\frac{2B m_q}{(4\pi f)^2}\log\frac{2Bm_q}{\mu^2}$ is replaced by $\xi\log\xi$, where the squared ratio of measured meson masses and decay constants $\xi=\frac{m_{\rm {PS}^2}}{(4\pi f_{\rm PS})^2}$ at the simulated quark mass $m_q$ is used.}%
with $(4\pi f_{\rm PS})$, effectively higher order contributions are resummed in an ad hoc manner. In the table I labeled this fit ansatz as ``$\xi$''$\chi$PT. For a detailed description of their fit functions and comparison to standard $\chi$PT see \cite{Noaki:2008iy}.

Until recently, $\chi$PT fits were only performed including the complete terms up to next-to-leading order (NLO) and leaving out higher order terms. In the typical range of pion masses up to 400 MeV, those omitted terms are usually estimated to have a 3--5\% effect. Therefore, if one aims for results with smaller systematic errors, one will ultimately have to include higher orders as well. One approach is to just add the analytic terms (which are multiplied by higher order LECs) to the fit formulae, ignoring the non-analytic (logarithmic) contributions. This has to be viewed as a (practical) phenomenological ansatz, but eventually it will be unsatisfactory if the predictions of $\chi$PT are to be tested by lattice simulations or one wants to measure accurately and in a well determined way  LECs up to a given order. For this one needs the complete formulae for the meson masses and decay constants in continuum (PQ)$\chi$PT up to that order. Bijnens et al.\ \cite{Bijnens:2004hk,Bijnens:2005ae,Bijnens:2005pa,Bijnens:2006jv}  published those up to NNLO and also provide \verb!FORTRAN! code for those fit functions upon request. The complete NNLO in $\chi$PT is now being used in the analyses by ETMC, MILC, JLQCD, and RBC-UKQCD (either for SU(2) and/or SU(3)). By including the complete NNLO the number of LECs which need to be fitted is increased by four in the case of SU(2) (two from the $\mathcal{O}(p^4)$ Lagrangian $\mathcal{L}_4$ and two from the $\mathcal{O}(p^6)$ Lagrangian $\mathcal{L}_6$) or in the case of SU(3) by ten (four from $\mathcal{L}_4$ and six from $\mathcal{L}_6$). In the partially quenched versions, there are 13 (5+8) or 15 (5+10) new LECs for PQ-SU(2) or SU(3), respectively. (For comparison: up to NLO there are only four (two from $\mathcal{L}_2$ and two from $\mathcal{L}_4$) in SU(2) or six (2+4) in PQ-SU(2), SU(3), and PQ-SU(3).) All groups report (independently), that at the moment their available data itself is not sufficiently sensitive to determine those additional LECs. In order to get meaningful fit results, currently all groups add priors or constraints for the new LECs from phenomenological estimates, which are available for the LECs originating from the $\mathcal{L}_4$ Lagrangian. For the remaining NNLO LECs, such estimates are not available. The sources for the phenomenological NLO LECs are the pion scalar radius, $\pi\pi$ scattering, pion charge radius or the axial form factor in $\pi\to l\nu\gamma$ in the case of SU(2) $\chi$PT. See \cite{Bijnens:2007yd} for more details and references. In the case of SU(3) $\chi$PT the best available estimates originate from combined phenomenological fits using the measured masses and decay constants of the pions and kaons plus data from $K_{l4}$ decays, cf.\ \cite{Bijnens:2007yd} and references therein. In general one should be cautious using such estimates as priors or constraints when the goal is to test the applicability of $\chi$PT by comparing it to lattice data. All those estimates had to be extracted under the assumptions that (NLO or NNLO) $\chi$PT \textit{is} sufficient to describe the experimental data available from the pion and kaon measurements, i.e., that $\chi$PT is valid at the pion and, more questionable, the kaon mass.

\begin{figure}
\begin{center}
\includegraphics[width=.45\textwidth]{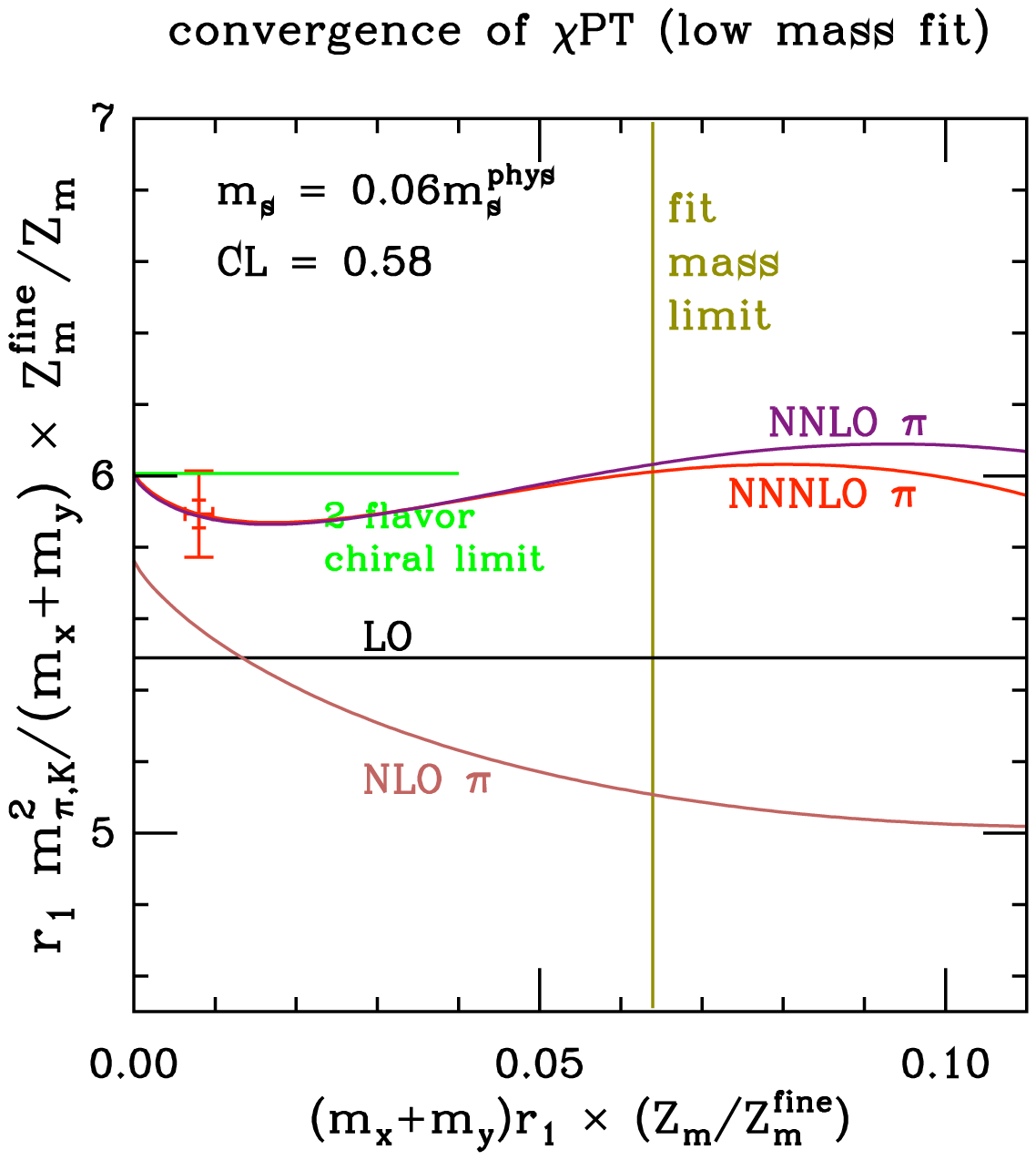}
\includegraphics[width=.48\textwidth]{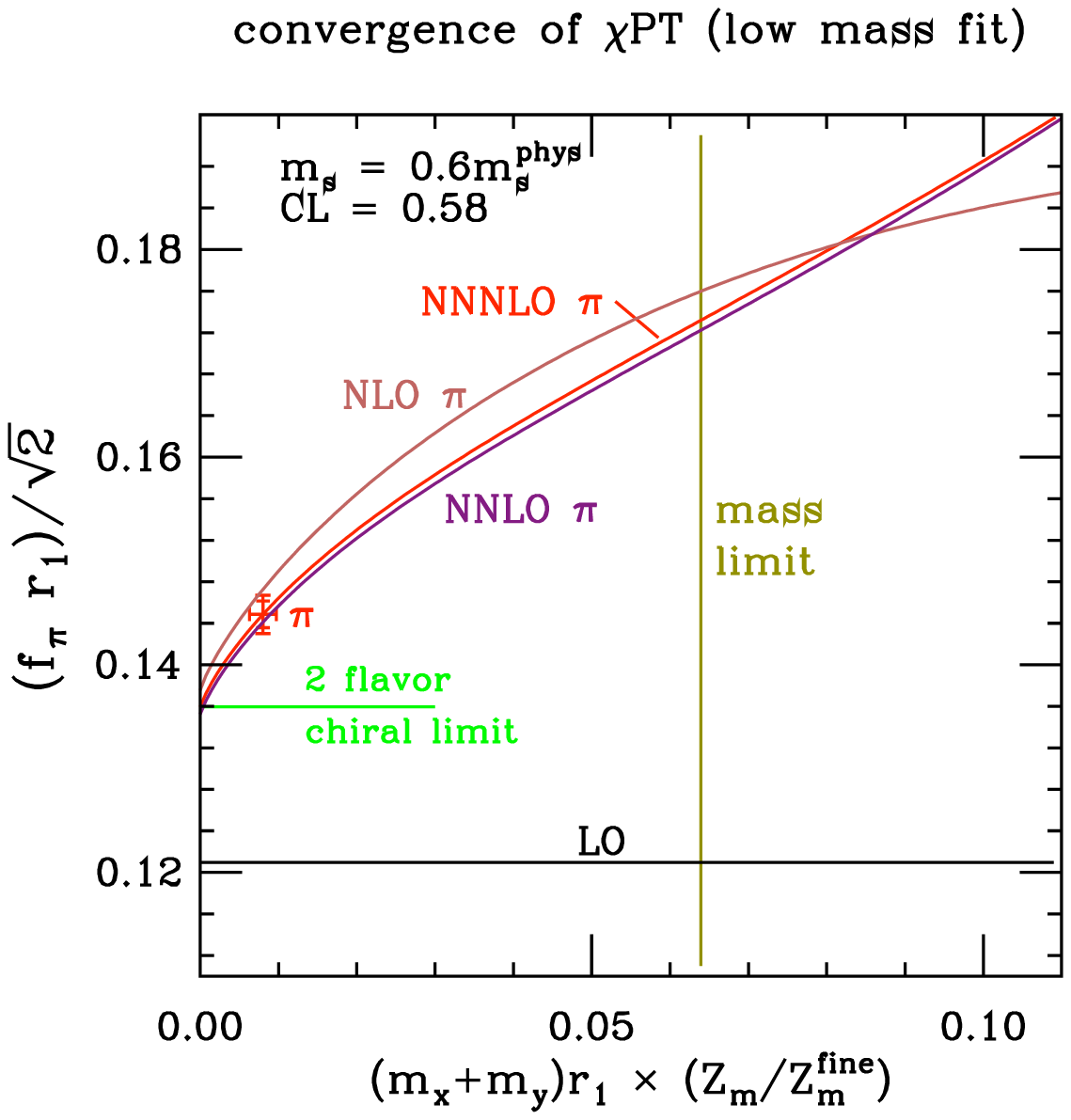}
\end{center}
\vspace*{\closercaption}
\caption{\label{fig:MILC_NNLO}SU(3) complete NNLO plus analytic N$^3$LO terms $\chi$PT fit for the squared meson masses \textit{(left panel)} and meson decay constants \textit{(right panel)} from the MILC-Collaboration \cite{Bazavov:2009fk,Bazavov:2009tw}, plots courtesy of C.~Bernard.}
\vspace*{\afterFigure}
\end{figure}

A remarkable observation in these NNLO $\chi$PT fits to lattice data is that for the meson masses the NNLO contribution is almost of the same magnitude as the NLO contribution, possibly indicating a poor convergence of the (asymptotic) series. This is illustrated in the left panel of Fig.~\ref{fig:MILC_NNLO}, which shows a NNLO fit performed by the MILC collaboration \cite{Bazavov:2009tw,Bazavov:2009ir}. ETMC \cite{Dimopoulos:2009hy} and RBC-UKQCD \cite{Mawhinney:2009jy} reported similar observations, when fitting their data with the complete NNLO. The anomalously small NLO contribution (being of the same size as the contribution from NNLO) has to be investigated further. Especially, it has to be excluded that this is influenced by the priors used in the fits. Given the fact that also a fit only using NLO describes the data well (with a bigger NLO contribution), one would expect NNLO to be a magnitude smaller. Comparing the extrapolations to either NLO or NNLO, it should be mentioned, that the extrapolated values are more or less consistent. Including NNLO terms therefore only shifts part of the NLO contribution to NNLO, when the latter is included. For the decay constants (shown in the right panel of Fig.~\ref{fig:MILC_NNLO}), a better convergence is observed. But here one should also keep in mind, that the NLO already describes a 20 or more per cent correction at NLO between the chiral limit and the physical point. The effect seen for the decay constant can be best described by saying that the addition of NNLO ``straightens'' the fit curve, making it look more linear in the region where data is available. Ultimately, simulations at lower meson masses will have to show, whether this linear trend continues or at which point the expected chiral curvature will appear. 

%
%

For completeness, it should be mentioned that as an alternative to the chiral fit formulae, also an analytic expansion (polynomial expansion) in the quark masses or meson masses around a non-zero point could be used. (This has been discussed under the name ``flavour expansion'' by L.~Lellouch in his Lattice 2008 plenary contribution \cite{Lellouch:2009fg}.) Results for a complete (meson masses and decay constants) analysis presented this year did not use an analytic expansion for their main results, which is why I will not discuss this point further in this context. 


\subsection{\label{subsec:chPTanalyses}$\chi$PT analyses of lattice data}

In the following, I will briefly summarize the main features of the chiral fits presented at this conference, before comparing the results of those fits in the remainder of this section. It should be mentioned, that all results discussed below either include finite volume effects in their analysis %
%
%
or their data has been corrected for these effects beforehand.

The \textit{MILC Collaboration}, using $N_f=2+1$ dynamical staggered fermions presented results from fits to SU(3) \cite{Bazavov:2009fk,Bazavov:2009tw} and SU(2) \cite{Bazavov:2009ir} rS$\chi$PT showing good agreement between those results. To have better control in their SU(3) extrapolation, also ``artificially light'' strange quarks are considered, which includes a $N_f=3$ simulation with three mass-degenerate light quarks as well. In a first step, they only fit their data for meson masses in the range of 180 to 380 MeV using complete NNLO (with priors for the LECs appearing only in NNLO terms). Since the inclusion of taste breaking effects is not available at complete NNLO, the root mean square averaged masses are used in NNLO, which is justified when taste breaking effects are negligible at that order due to a fine enough lattice spacing. In a second step, where the LECs obtained in the first step are now fixed, the fit range is enlarged to meson masses up to 550 MeV (to include the physical kaon mass) and analytic-only N$^3$LO and N$^4$LO terms are added, see Fig.~\ref{fig:MILC_NNLO} for an example fit. Currently two different lattice spacings are used in the first step (0.09 fm, 0.06 fm) of the analysis and in the second step simulations at an even finer lattice spacing (0.045 fm) are considered as well before the continuum limit is taken.

The \textit{RBC-UKQCD Collaborations} fit their data obtained with $N_f=2+1$ domain wall fermions to SU(2) $\chi$PT up to NLO in a range of 290 to 420 MeV for the light pseudo-scalar meson \cite{Kelly:2009fp}. The light-strange meson (i.e. the kaon if the light quark mass approaches the physical value of $m_{\rm ud}$) is described in SU(2) $\chi$PT as well. Data obtained at two different lattice spacings (0.09 fm, 0.12fm) have been used and a continuum extrapolation (assuming $\mathcal{O}(a^2)$ scaling) is included in the analysis. This is a continuation of their previous work \cite{Allton:2008pn}, where only the coarser of the two lattice spacings was available. The simulations are done at a fixed value for the strange quark, which is reweighted during the analysis to fine-tune to its physical value. To estimate the systematic effect of neglected higher orders in $\chi$PT, a comparison with results from an analytic expansion is performed. First results are also available from fits using complete NNLO $\chi$PT \cite{Mawhinney:2009jy}, but currently those are not used for the final quoted values. 

The \textit{PACS-CS Collaboration} presented a detailed comparison of NLO SU(2) and SU(3) $\chi$PT fits to their $N_f=2+1$ improved Wilson fermions data at a single lattice spacing at last year's conference \cite{Kuramashi:2008tb} and published those as well \cite{Aoki:2008sm}. Both, SU(2) and SU(3) $\chi$PT have been considered and a similar observation to that from RBC-UKQCD \cite{Allton:2008pn} about the better convergence of SU(2) $\chi$PT has been obtained. As already mentioned above, for their current analysis they pursue a different strategy, namely to reweight their lightest simulated point at a meson mass of approximately 160 MeV to the physical pion mass \cite{Kuramashi:LAT09}. 

The \textit{JLQCD Collaboration} now also has data available from $N_f=2+1$ dynamical overlap fermion simulations \cite{Noaki:LAT09} in addition to their previous simulation at $N_f=2$ \cite{Noaki:2008iy}, each at a single lattice spacing of 0.10 fm and 0.12 fm, respectively. They perform the extrapolation from their mass range of 320 to 800 MeV to the physical point by using NNLO ``$\xi$'' SU(2) and SU(3) $\chi$PT (see explanation above), since in \cite{Noaki:2008iy} they argued that this gives the best description of their data.

The \textit{ETM Collaboration} presented results from their simulations with 2 flavors of twisted mass fermions \cite{Dimopoulos:2009hy,ETMC:chPTprep}. (Preliminary results for 2+1+1 flavors have been presented as well \cite{Baron:2009zq}.) For their meson masses in the range of 250 to 600 MeV they use complete SU(2) $\chi$PT up to NNLO (including priors for NNLO LECs). The continuum extrapolation is taken from three different lattice spacings (0.05, 0.07, and 0.08 fm). To obtain their final quoted values and the error estimate, they perform several different fits (varying fit ranges, using either NLO or NNLO,\ldots) and finally average the results weighted by the quality of the fit. The strange quark is quenched in their main analysis and the kaon decay constant and mass (or the thereby extracted mass of the strange quark) are obtained in a partially quenched set-up \cite{Blossier:2009bx,Tarantino:private}. 

In addition, the overview table also contains the mixed action results presented already at last year's conference by \textit{Aubin, Laiho, Van de Water}\cite{Aubin:2008ie} measured with domain wall fermions on configurations generated with $N_f=2+1$ dynamical staggered quarks by MILC. A recent update of their work has been presented elsewhere \cite{Laiho:CD09}. Currently two lattice spacing (0.12 fm, 0.09 fm) are used for the continuum extrapolation. Complete SU(3) MA$\chi$PT is used up to NLO, but higher-order analytic-only terms had to be added to obtain reasonable fit results.

\subsection{\label{subsec:chPTresults}Results}
\subsubsection{\label{subsubsec:chPTresults:Decay}Decay constants}

In Table~\ref{tab:chPTdecays}, I compiled the results for the pion and kaon decay constants as well as the meson decay constant in the chiral limit of SU(2) and SU(3) $\chi$PT, $f$ and $f_0$, respectively. Two collaborations use the pion decay constant to set the lattice scale, so no prediction for $f_\pi$ is available in those cases. With exception of JLQCD $N_f=2$, all the quoted values agree within errors with the experimentally measured value \cite{Amsler:2008zzb}, although it has to be noted that in some cases the combined statistical and systematic error is as big as 7 per cent. The same picture emerges for the kaon decay constant, although combined errors here are 4 per cent at most.%
\footnote{It is also remarkable to note that the value quoted by PDG in 2008 \cite{Amsler:2008zzb} shifted by roughly $2.5\sigma$ compared to the previously quoted value in 2006 \cite{Yao:2006px}.}%
\newcounter{fKfPiFootnote}
\setcounter{fKfPiFootnote}{\value{footnote}}

\begin{figure}
\begin{minipage}{.55\textwidth}
\begin{center}
\includegraphics[width=.95\textwidth]{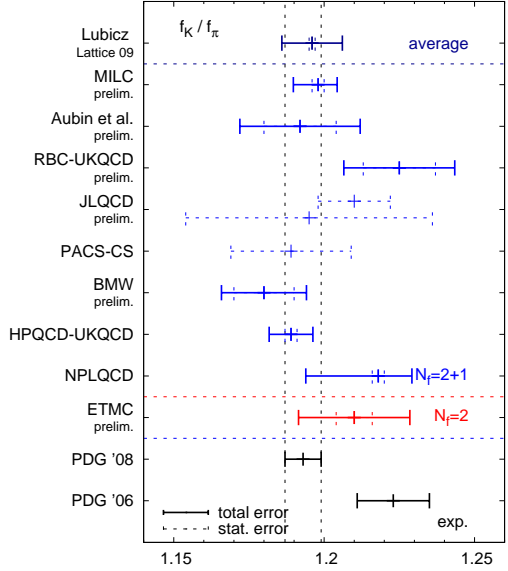}
\end{center}
\end{minipage}
\begin{minipage}{.45\textwidth}
\caption{\label{fig:fKfPi} The ratio of the kaon and pion decay constants $f_K/f_\pi$ from lattice simulations (for details see text) compared to the experimentally measured value (PDG '08) \cite{Amsler:2008zzb}. Also shown is the PDG '06 value \cite{Yao:2006px}, see remark in footnote \arabic{fKfPiFootnote}. Other values taken from: Lubicz \cite{Lubicz:2010nx}, MILC \cite{Bazavov:2009tw}, Aubin et al.\ \cite{Laiho:CD09}, RBC-UKQCD \cite{Kelly:2009fp,Mawhinney:2009jy}, JLQCD \cite{Noaki:LAT09,Noaki:private09}, PACS-CS \cite{Aoki:2008sm}, BMW \cite{Ramos:2010ar}, HPQCD-UKQCD \cite{Follana:2007uv}, NPLQCD \cite{Beane:2006kx}, ETMC \cite{Dimopoulos:2009hy,Blossier:2009bx}.}
\end{minipage}
\vspace*{\afterFigure}
\end{figure}

Figure~\ref{fig:fKfPi} shows the ratio of decay constants $f_K/f_\pi$ compared to the experimentally observed value \cite{Amsler:2008zzb}. In addition, preliminary results from the BMW Collaboration \cite{Ramos:2010ar} and older results from NPLQCD \cite{Beane:2006kx} and HPQCD \cite{Follana:2007uv} have been included as well. While the BMW Collaboration performed a direct extrapolation of the ratio (instead of separately extrapolating $f_\pi$ and $f_K$), NPLQCD combined an SU(3) extrapolation of $f_K$ with the experimentally measured value of $f_\pi$, and HPQCD used chiral expansions with priors. An updated average for the ratio of decay constants was given by V.~Lubicz at this conference \cite{Lubicz:2010nx}: $f_K/f_\pi=1.196(1)(10)$, which is especially interesting for the determination of the CKM matrix element ratio $|V_{\rm ud}|/|V_{\rm us}|$, see the contribution of R.S.~Van~de~Water \cite{VandeWater:2009uc} and also \cite{Sachrajda:2009ma} for more details and other implications in CKM physics.  

%
%
\begin{table}
\begin{center}
\begin{tabular}{rcccc}
\hline\hline
      & $f_\pi\:[{\rm MeV}]$ & $f_K\:[{\rm MeV}]$ & $f\:[{\rm MeV}]$ & $f_0\:[{\rm MeV}]$ \\\hline
ETMC  & input & 158.1(0.8)(2.0)(1.1) & {\it 121.57(70)} &  \\
JLQCD (2) & \multicolumn{2}{l}{119.6(3.0)(1.0)$(^{+6.4}_{-0.0})$}  & \multicolumn{2}{l}{111.7(3.5)(1.0)$(^{+6.0}_{-0.0})$} \\
JLQCD (2+1) & input & {\it 157.3(5.5)} & {\it 121(14)} & {\it 79(20)} \\
RBC-UKQCD  & {\it 122.2(3.4)(7.3)} & {\it 149.7(3.8)(2.0)} & {\it 113.0(3.8)(6.8)} &  93.5(7.3) \\
PACS-CS  & 134.0(4.3) & 159.4(3.1) & 126.4(4.7) & 118.5(9.0)  \\
MILC  & {\it 128.0(.3)(2.9)} & {\it 153.8(0.3)(3.9)} & {\it 122.8(.3)(.5)} & {\it 111.0(2.0)(4.1)} \\ 
Aubin et al. & {\it 131.1(1.3)(2.2)} & {\it 156.3(1.3)(2.0)}  \\\hline
PDG    & 130.4(.04)(.2) & 155.5(.2)(.8)(.2) & -- & --  
\\\hline\hline
\end{tabular}
\end{center}
\caption{\label{tab:chPTdecays}Results for the pion ($f_\pi$) and kaon ($f_K$) decay constants from the chiral fits and their experimentally measured values \cite{Amsler:2008zzb}, see also footnote \arabic{fKfPiFootnote}. Also given are the decay constants in the chiral limit of SU(2) ($f$) and SU(3) $\chi$PT ($f_0$). Results in \textit{italics} mark preliminary results. In the case of the JLQCD collaboration results from their earlier 2 \cite{Noaki:2008iy} and recent 2+1 \cite{Noaki:LAT09,Noaki:private09} flavor simulations are shown. The RBC-UKQCD results are from \cite{Kelly:2009fp,Mawhinney:2009jy} except for $f_0$, which is is from their earlier work \cite{Allton:2008pn}. Other results taken from: ETMC \cite{Dimopoulos:2009hy, Blossier:2009bx}, PACS-CS \cite{Aoki:2008sm}, MILC  \cite{Bazavov:2009tw}, Aubin et al.\  \cite{Laiho:CD09}.}
\end{table}


%
%

\subsubsection{\label{subsubsec:results:LECs}Low energy constants}

\begin{figure}
\begin{center}
\includegraphics[width=.475\textwidth]{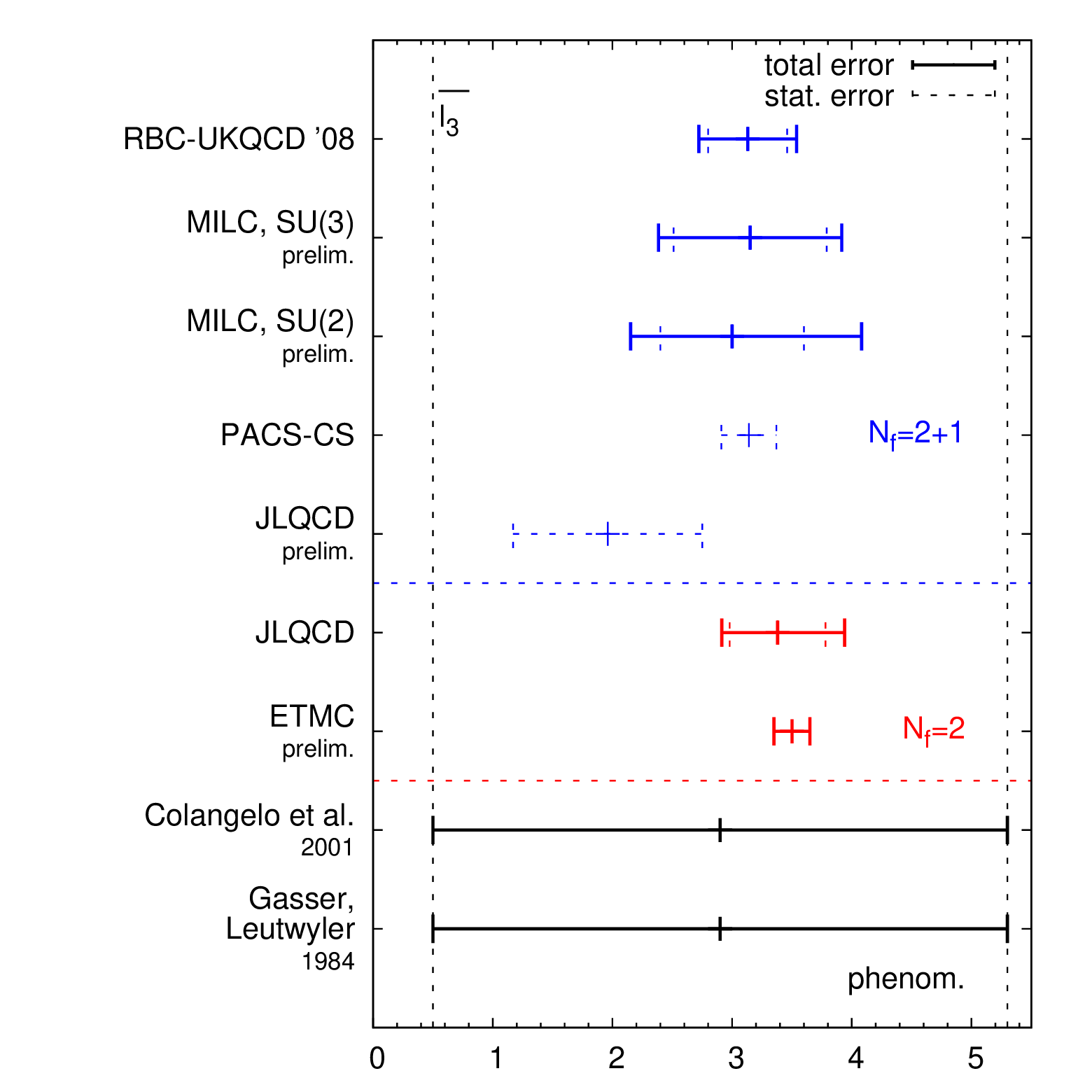}
\includegraphics[width=.475\textwidth]{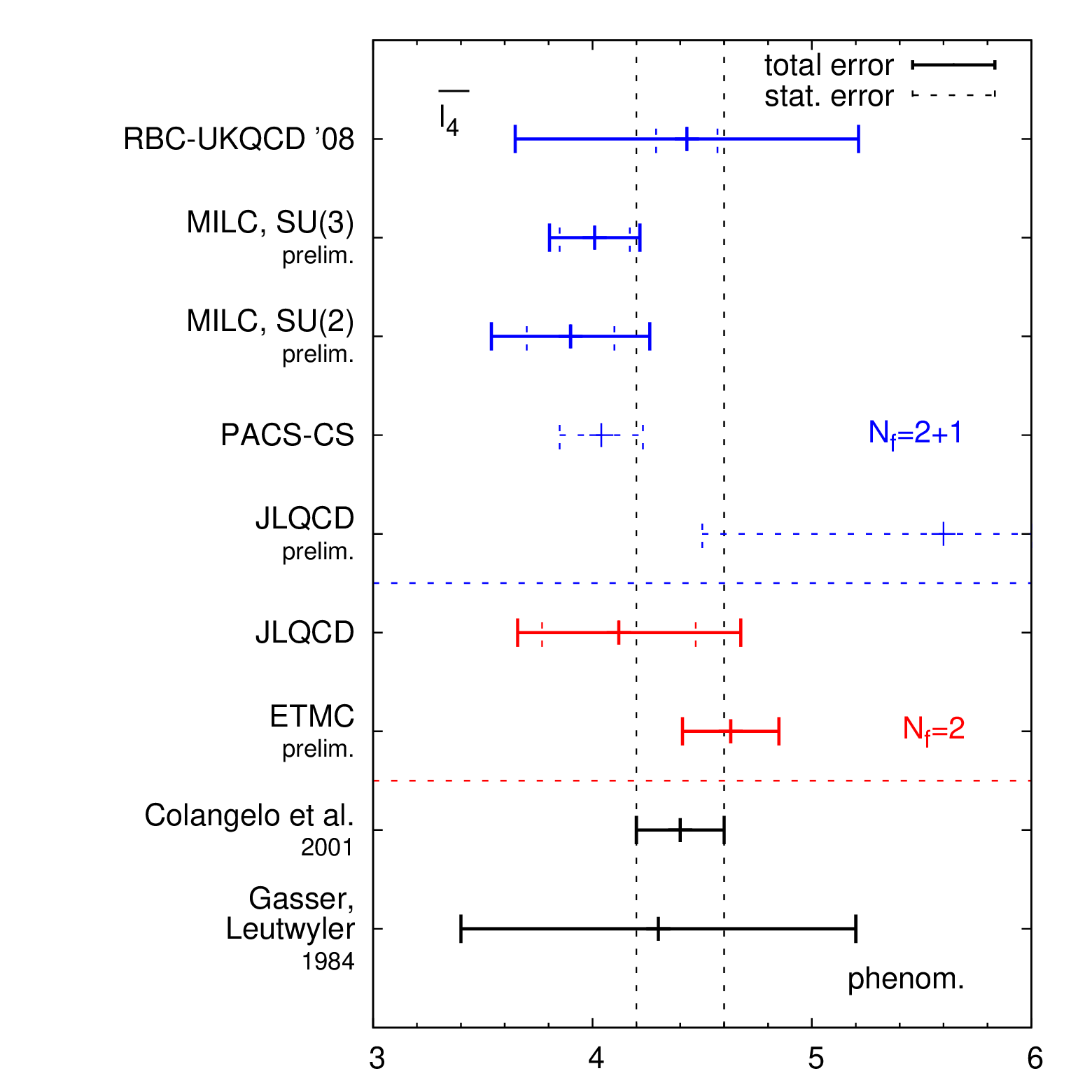}
\end{center}
\vspace*{\closercaption}
\caption{\label{fig:lbar34}The LECs $\bar{l}_3$ \textit{(left panel)} and $\bar{l}_4$ \textit{(right panel)} in SU(2) $\chi$PT (defined at a scale $m_\pi$) from different lattice simulations, compared to phenomenological estimates \cite{Gasser:1983yg,Colangelo:2001df}, cf.\ also \cite{Bijnens:2007yd}. The RBC-UKQCD result is from their earlier work \cite{Allton:2008pn}, other values are taken from: MILC \cite{Bazavov:2009tw,Bazavov:2009ir}, PACS-CS \cite{Aoki:2008sm}, JLQCD \cite{Noaki:LAT09,Noaki:private09}, ETMC \cite{Dimopoulos:2009hy,ETMC:chPTprep}.}
\vspace*{\afterFigure}
\end{figure}

The results for the SU(2) LECs $\bar{l}_3$ and $\bar{l}_4$ are shown in Fig.~\ref{fig:lbar34}. Those are of interest in phenomenological applications, e.g., the pion-pion scattering length, see \cite{Leutwyler:2008xd}.  The LECs are as usual defined at the scale of the physical (charged) pion mass. Within the quoted uncertainties, no distinction can be made between LECs from $N_f=2+1$ simulations, which include the effects of the strange quark, and those which only use $N_f=2$ and therefore do not account for the strange-quark effects. The lattice simulations confirm the phenomenological estimates \cite{Gasser:1983yg,Colangelo:2001df,Bijnens:2007yd} and in the case of $\bar{l}_3$ are also able to provide a value with smaller uncertainty. An interesting remark should be made at this point: the phenomenological estimate of $\bar{l}_4=4.4\pm0.2$ \cite{Colangelo:2001df,Bijnens:2007yd} together with the requirement that SU(2) $\chi$PT at NLO returns the experimentally measured value for $f_\pi$ constrains the decay constant in the SU(2) chiral limit to $f\approx (121.5\pm1.0)\,{\rm MeV}$.

%
%

\subsubsection{\label{subsubsec:results:quarkmasses}Quark masses}

Since $\chi$PT describes the dependence of the meson quantities on the quark masses, it allows one to extract the (light) quark masses once a reliable fit has been achieved. Usually the (experimentally observed) neutral pion mass is used to define the point of the average up/down quark mass $m_{\rm ud}=(m_{\rm u}+m_{\rm d})/2$ and the kaon mass to define the strange quark mass.%
\footnote{This is a somewhat simplified statement given the complexity of what is nowadays standard in $\chi$PT fits to lattice data. E.g., in a complete analysis, one in general needs three input parameters to fix $m_{\rm ud}$, $m_{\rm s}$, and the lattice scale $1/a$. Commonly, the pion and kaon masses are used plus a third quantity like $r_0$, $r_1$, $m_\Omega$ or $f_\pi$. In the case the third quantity depends on $m_{\rm ud}$ and/or $m_{\rm s}$ (e.g.\ $f_\pi$ or $m_\Omega$) a global fit procedure has to be carried out. But still the quark masses are mainly influenced by the input meson masses, so that the simplified statement is justified.}%
\ The quark masses which enter a lattice simulation are bare parameters defined in the lattice regularization scheme and depend on the fermion action and lattice scale. Therefore, the masses have to be renormalized, commonly the $\overline{\rm MS}$ scheme at a renormalization scale of $\mu=2\,{\rm GeV}$ is chosen. This transformation can either be performed perturbatively (up to some given order) or non-perturbatively by measuring the renormalization factors for the conversion to a regularization independent (RI) scheme for specific operators (taking into account possible operator mixing) directly on the lattice. In the latter case, the conversion from the RI to the $\overline{\rm MS}$ scheme still has to be performed perturbatively. For more details on the renormalization of quark masses and operators in lattice simulations, see the plenary contribution by Y.~Aoki at this conference \cite{Aoki:2010yq}. In the following, I will quote quark mass results in the $\overline{\rm MS}(\mu=2\,{\rm GeV})$ scheme and indicate whether non-perturbative or perturbative renormalization techniques have been used.

\begin{figure}
\begin{minipage}{.65\textwidth}
\begin{center}
\includegraphics[width=\textwidth]{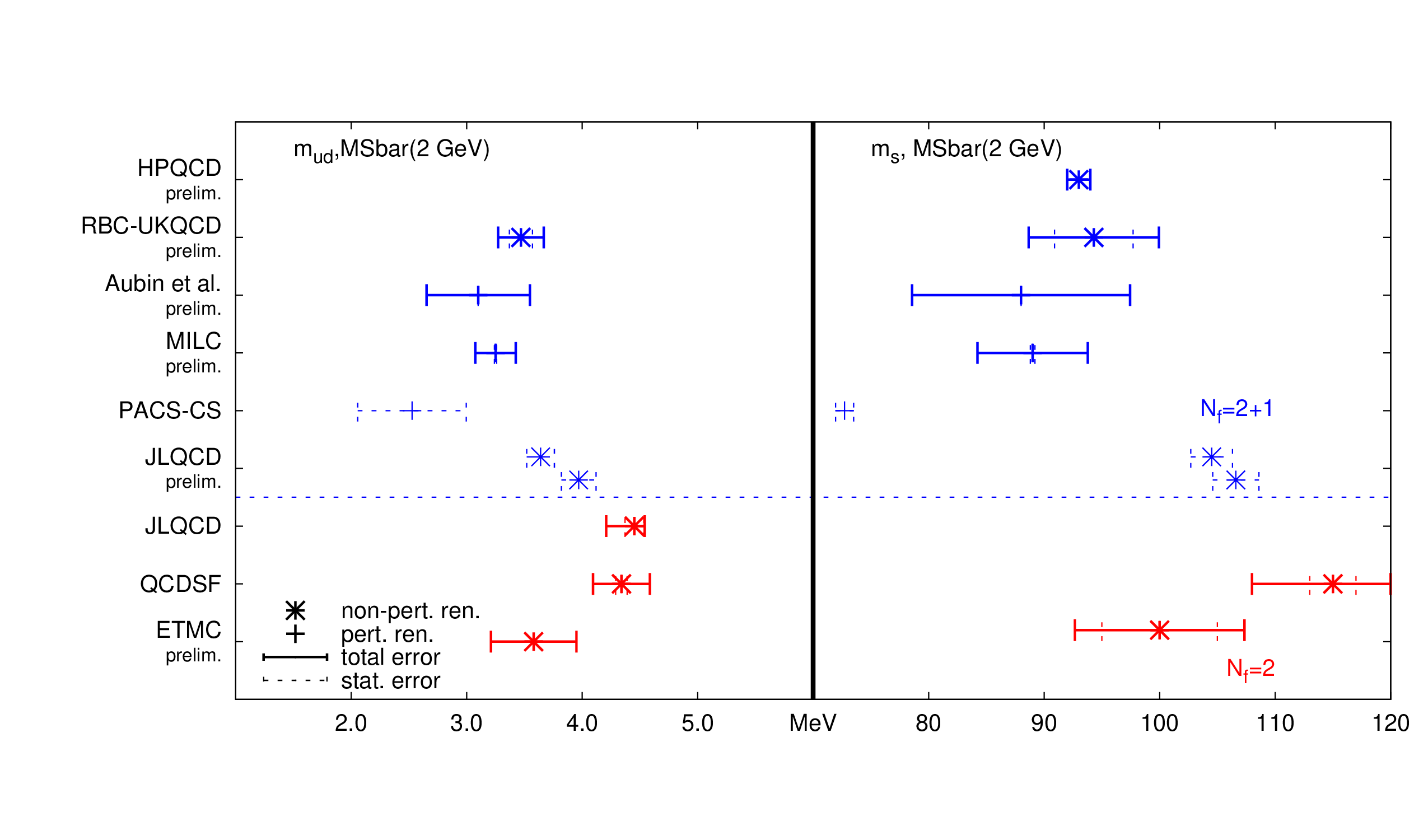}
\end{center}
\end{minipage}
\begin{minipage}{.35\textwidth}
\begin{center}
\includegraphics[width=\textwidth]{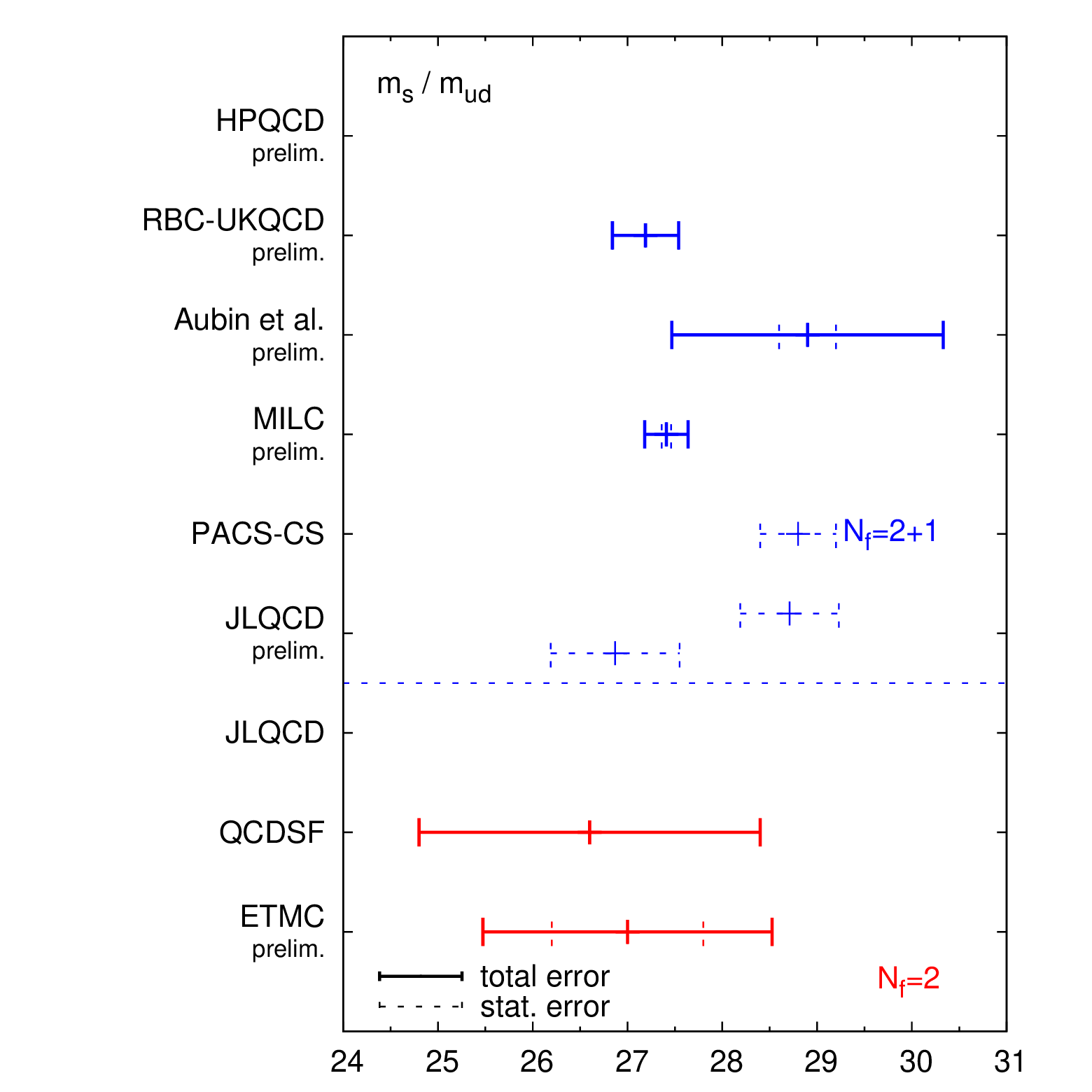}
\end{center}
\end{minipage}
\caption{\label{fig:quarkMasses}\textit{Left panel:} The quark masses $m_{\rm ud}$ and $m_{\rm s}$ renormalized in the $\overline{\rm MS}$-scheme at $\mu=2\,{\rm GeV}$. \textit{Right panel:} The quark mass ratio $m_{\rm s}/m_{\rm ud}$. For details see text, values taken from HPQCD \cite{McNeile:2009eq}, RBC-UKQCD \cite{Kelly:2009fp,Mawhinney:2009jy}, Aubin et al.\ \cite{Laiho:CD09}, MILC \cite{Bazavov:2009tw}, PACS-CS \cite{Aoki:2008sm}, JLQCD \cite{Noaki:LAT09,Noaki:private09,Noaki:2008iy}, QCDSF \cite{Gockeler:2006vi}, ETMC \cite{Dimopoulos:2009hy,Tarantino:private}.}
\vspace*{\afterFigure}
\end{figure}

Figure~\ref{fig:quarkMasses} shows a compilation of both $m_{\rm ud}$ and $m_{\rm s}$ quark masses and their ratio obtained from the analyses discussed above. Also included are previous results from the PACS-CS \cite{Aoki:2008sm}, JLQCD ($N_f=2$) \cite{Noaki:2008iy}, and QCDSF \cite{Gockeler:2006vi} collaborations, as well as the HPQCD result presented at this conference \cite{McNeile:2009eq}. The latter work used a different approach, namely to extract the mass of the strange quark from the strange/charm quark mass ratio $m_{\rm s}/m_{\rm c}$. Excluding the PACS-CS and preliminary JLQCD $N_f=2+1$ points, which currently do not provide an estimate for their systematic uncertainty, the data might show a slight trend to higher quark masses observed in $N_f=2$ simulations (leaving out the effect of a dynamical strange quark) compared to $N_f=2+1$ simulations. But for a definite statement, the (mainly) systematic uncertainties have to be reduced further. 

%
%

The mass splitting between the up and down quarks can be estimated, e.g., by incorporating electro-magnetic effects by Dashen's theorem and the violation of the latter, and estimating the ratio $m_{\rm u}/m_{\rm ud}$ from the observed mass difference between the neutral and charged kaons, see e.g.\ \cite{Aubin:2004fs}. Results have been presented by the MILC Collaboration \cite{Bazavov:2009tw} and Aubin et al.\ \cite{Laiho:CD09}, see Tab.~\ref{tab:quarkMassSplit} for a summary of their preliminary results. The RBC-UKQCD Collaboration presented preliminary results for the mass splittings at this conference, where the electro-magnetic effects have been included in the lattice measurements of $N_f=2$ and $N_f=2+1$ QCD via quenched QED \cite{Zhou:2009ku}.

\begin{table}
\begin{center}
\begin{tabular}{lccc}
\hline\hline
             & $m_{\rm u}\:[{\rm MeV}]$ & $m_{\rm d}\:[{\rm MeV}]$ & $m_{\rm u}/m_{\rm d}$ \\\hline
MILC         & 1.96(0)(6)(10)(12)      & 4.53(1)(8)(23)(12)      & 0.432(1)(9)(0)(39) \\
Aubin et al. & 1.7(0)(2)(2)(1)         & 4.4(0)(2)(4)(1)         & 0.39(1)(3)(0)(4)\\\hline\hline
\end{tabular}
\end{center}
\caption{\label{tab:quarkMassSplit}The up and down quark masses $m_{\rm u}$, $m_{\rm d}$ ($\overline{\rm MS}$-scheme, $\mu=2\,{\rm GeV}$) and their ratio from \cite{Bazavov:2009tw,Laiho:CD09} (preliminary results).}
\vspace*{\afterTable}
\end{table}

\section{\label{sec:hadronSpect}Hadron spectrum from lattice QCD}

%
%
\begin{figure}
\begin{minipage}{.675\textwidth}
\begin{center}
\includegraphics[width=.9\textwidth]{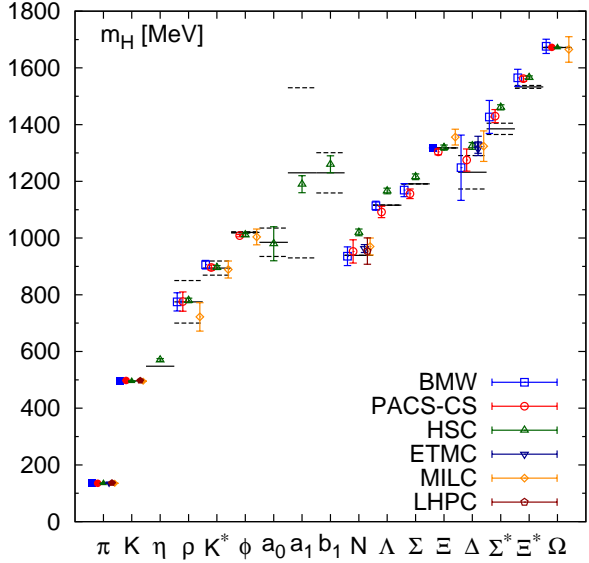}
\end{center}
\end{minipage}
\begin{minipage}{.3\textwidth}
\caption{\label{fig:hadronSpect}The mass spectrum of the light hadrons from dynamical lattice QCD simulations (from \cite{Durr:2008zz,Aoki:2008sm,Lin:2008pr,Alexandrou:2008tn,Bazavov:2009bb,Gottlieb:private,WalkerLoud:2008bp}, see text for details), compared to the masses and widths (\textit{solid} and \textit{dashed} lines, resp.) observed in Nature \cite{Amsler:2008zzb}.}
\end{minipage}
\vspace*{\afterFigure}
\end{figure}

After limiting the discussion to the light pseudo-scalar sector in the previous section, I will now include light vector and scalar mesons and light baryons as well, turning to the complete light hadron spectrum. The mass spectrum of the light hadrons has been extracted from lattice simulations by various groups up to now, providing a good cross-check between the different lattice fermion formulations, experimental inputs and extrapolation methods used. In Fig.~\ref{fig:hadronSpect} I show a compilation of some recent results for the light meson masses and octet and decuplet baryon masses from lattice simulations compared with the masses (and widths) observed in Nature \cite{Amsler:2008zzb}. Obviously, the lattice simulation results are consistent among each other and reproduce the experimentally measured values. As already discussed in the previous section, to fix the quark masses ($m_{\rm ud}$ and $m_{\rm s}$ or just $m_{\rm ud}$ if only non-strange hadrons are considered) and the overall lattice scale, one needs three (or two in the case of non-strange hadrons only) input parameters. In all the analyses discussed here, the pion and kaon masses have been used for the quark masses, while choices for the third quantity may be the mass of the nucleon, the $\Omega$, or the $\Xi$, the pion decay constant or the quark potential ($r_0$, $r_1$).

In the following, I will highlight some details of the analyses leading to the results shown in Fig.~\ref{fig:hadronSpect}. The \textit{Budapest-Marseille-Wuppertal (BMW) Collaboration} \cite{Durr:2008zz} calculated the spectrum from their 2+1 flavor simulation using improved Wilson fermions at three different values for the lattice spacing and used a polynomial extrapolation from their simulated lightest meson masses in the range of 190 to 650 MeV down to the physical point. The \textit{Hadron Spectrum Collaboration (HSC)} presented first results from a study performed to tune the mass of the strange quark in their simulations \cite{Lin:2008pr,Cohen:2009zk}. Currently, they have only one lattice spacing available using 2+1 flavors of anisotropic clover fermions. Also a polynomial expansion is used to extrapolate from their lightest meson masses at  370 -- 1520 MeV down to the physical pion mass. The \textit{PACS-CS Collaboration} has data at a single value for the lattice spacing with 2+1 flavors of improved Wilson fermions available very close to the physical point, namely light meson masses  ranging from 160 to 700 MeV. In their initial study \cite{Aoki:2008sm} they used a (short) polynomial extrapolation to the physical point, whereas at this year's conference preliminary results obtained by directly reweighting to the physical pion mass have been presented \cite{Kuramashi:LAT09}. The \textit{Lattice Hadron Physics Collaboration (LHPC)} is using a mixed action approach to study the hadron spectrum. Domain wall valence propagators are calculated on the 2+1 dynamical flavors of improved staggered fermions at a single lattice spacing with lightest meson masses of 300 MeV and above \cite{WalkerLoud:2008bp}. They studied various different ans\"atze for the extrapolation towards the physical point, cf.\ also \cite{WalkerLoud:2008pj} and the discussion in Sec.~\ref{subsec:baryonExtr}. The \textit{MILC Collaboration} has results available from 2+1 flavor simulations with improved staggered fermions, currently including three different values for the lattice spacings and a lightest meson with a mass of 180 MeV. They use, depending on the quantity, either chiral or polynomial extrapolations \cite{Bazavov:2009bb}. The \textit{European Twisted Mass Collaboration (ETMC)} calculated the nucleon and $\Delta$ baryon masses (i.e.\ only non-strange quantities) in their set-up of 2 flavor twisted mass fermions \cite{Alexandrou:2008tn,Alexandrou:2009ng} from three lattice spacings at light meson masses down to 270 MeV using a chiral extrapolation.

In addition to the above results, which are included in the summary plot (Fig.~\ref{fig:hadronSpect}), the \linebreak RBC-UKQCD Collaborations and the QCDSF-UKQCD Collaboration also published (preliminary) results for the nucleon mass. The \textit{RBC-UKQCD Collaborations} showed preliminary results for the nucleon mass extrapolation from two values of the lattice spacing with domain wall fermions \cite{Maynard:2010wg}. The current work of the \textit{QCDSF-UKQCD Collaboration} focuses on the splitting of the octet and decuplet baryon masses in the case of SU(3) symmetry breaking with 2+1 flavors of improved Wilson (``SLiNC'') fermions \cite{Bietenholz:2009fi} and the study of the $\rho$ and $\Delta$ resonances with 2 flavors of clover fermions \cite{Schierholz:LAT09}, see Sec.~\ref{subsec:rho}, where the nucleon mass has been used to set the lattice scale.

\subsection{\label{subsec:baryonExtr}Extrapolations for baryon masses}

As for the meson sector, the formulae for the baryon mass extrapolation to the physical point can be based on chiral symmetry arguments. One approach is heavy baryon chiral perturbation theory (HB$\chi$PT) \cite{Jenkins:1990jv,Bernard:1992qa}, where the effective fields again are either the pions in SU(2) HB$\chi$PT or the pions, kaons, and the $\eta$ in SU(3) HB$\chi$PT. In general, the quark mass dependence of a baryon mass $M_{\rm baryon}$ in HB$\chi$PT reads
\[ M_{\rm baryon} \;=\; M^{(0)}_{\rm baryon}\:+\:M^{(1)}_{\rm baryon}\:+\:M^{(3/2)}_{\rm baryon}\:+\:\ldots\,, \]
where every term shows a scaling with the quark mass $m_q$ according to
\[ M_{\rm baryon}^{(i)}\;\propto\;m_q^i\,. \]
Therefore, the expansion parameter here is $\epsilon\simeq m_{\pi(K,\eta)}/\Lambda_\chi$ rather than $\simeq m_{\pi(,K,\eta)}^2/\Lambda_\chi^2$ as in meson $\chi$PT (but in both cases, the NLO term is of order $m_{\pi,(K,\eta)}^2/\Lambda_\chi^2$). Alternatively, analytic expansions around the physical point or the chiral limit are used for the extrapolation, cf.\ also the review on this subject given by A.~Walker-Loud at last year's conference \cite{WalkerLoud:2008pj}. In the following, I will review the current status of baryon mass extrapolations done by the groups mentioned above and also review work being done to study the SU(3)-breaking effects in baryon masses.

SU(2) HB$\chi$PT has been studied with 2+1 dynamical fermion flavors by PACS-CS (improved Wilson fermions) \cite{Aoki:2008sm} and the LHP Collaboration (mixed action: dynamical improved Wilson with valence domain wall fermions) \cite{WalkerLoud:2008bp} and with 2 dynamical fermion flavors by ETMC (Wilson twisted mass) \cite{Alexandrou:2008tn,Alexandrou:2009ng} and QCDSF (clover-Wilson fermions) \cite{Schierholz:LAT09}. While the latter two claim to observe a good agreement between their data and the predictions of SU(2) HB$\chi$PT, the PACS-CS Collaboration reports the theory to have a small convergence radius and an extrapolation which misses the physical point. LHPC finds that the fits describe their data adequately, but the extracted axial and nucleon-$\Delta$ couplings from those fits are inconsistent with phenomenological expectations. It should be mentioned, that the LHPC data has not been corrected for finite size effects, which seem to have an important impact in the other analyses.

Fitting their measured lattice data to the predictions of SU(3) HB$\chi$PT has been pursued by PACS-CS and LHPC. Both LO and NLO fits are possible, but again discrepancies are observed between the octet and octet-decuplet axial couplings (commonly referred to as $\mathcal{D}$, $\mathcal{F}$, and $\mathcal{C}$) obtained from the fits and phenomenological models or direct lattice calculations of these couplings.

Analytic expansions for the baryon masses were used in the analyses of BMW \cite{Durr:2008zz}, PACS-CS \cite{Aoki:2008sm}, and HSC \cite{Lin:2008pr,Cohen:2009zk} and also in part in MILC's analysis. Those seem to work fine, if the available data is close enough to the physical point as can be confirmed by simulations performed directly at the physical point (or reweighted to this point, as has been presented in \cite{Kuramashi:LAT09}). But one has to keep mind, that now the simulated lattice volume has to be large enough to exclude finite size effects since currently all methods to correct for such effects in a volume that is too small rely on (NLO)$\chi$PT.

In a recent publication, Jenkins et al.\ \cite{Jenkins:2009wv} studied the effects of SU(3)-breaking via baryon mass relations, which they compared to actual data from lattice simulations. Using the $1/N_c$ expansion, one can establish such mass relations which are expected to be fulfilled at $\mathcal{O}(N_c)$ leading to effects of the order of 1300 MeV, and subsequently $\mathcal{O}(1)$ or $\mathcal{O}(1/N_c)$ relations leading to effects of the order of 430 MeV or 140 MeV, respectively. By this method, it is possible to study the validity of the baryon mass relations and in turn the applicability of HB$\chi$PT in a systematic way. Those mass relations turn out to be valid in the expected range, although the same mismatch for the couplings in HB$\chi$PT compared with phenomenological estimates is observed. The QCDSF Collaboration in their current $N_f=2+1$ dynamical clover-Wilson simulations also studies the effects of SU(3) breaking \cite{Bietenholz:2009fi}. They simulate at different sets of two light and one heavier quark flavor with the constraint of keeping $2m_l+m_h$ constant, including the point $m_l=m_h$.

To conclude, the status of extrapolations using HB$\chi$PT is somewhat unsatisfactory at the moment, at least if SU(3) HB$\chi$PT is considered. Further investigations, possibly including NNLO and artificially light strange quarks could reveal an answer for the convergence radius and therefore if the strange quark mass lies within this region. With the data available at this time, short polynomial extrapolations or reweighting techniques seem to be more successful to extract the baryon spectrum. Also it should be mentioned%
\footnote{I am thankful to Thomas R.\ Hemmert for pointing this out to me.}
 that relativistic (also called covariant) baryon $\chi$PT (B$\chi$PT) \cite{Gasser:1987rb,Becher:1999he}  is an alternative approach based on chiral effective Lagrangians \cite{Bernard:2003rp,Procura:2003ig}. The QCDSF Collaboration studied, e.g., the nucleon mass behavior within B$\chi$PT including finite volume effects \cite{AliKhan:2003cu,Dorati:2007bk}. Also the ETM Collaboration compared their SU(2) HB$\chi$PT fits mentioned above with fits based on the predictions from relativistic B$\chi$PT, stating good agreement between the two \cite{Alexandrou:2008tn}.

\subsection{\label{subsec:rho}$\rho$ vector meson mass and decay constant}

The lightest vector meson, the $\rho$, is unstable in Nature, since the decay into two pions is allowed. In lattice simulations, where the pion masses are light enough ($2m_\pi<m_\rho$), the $\rho$ therefore has to be treated as a resonance. Its mass can be calculated from the finite volume dependence of the phase shift of the $\pi\pi$ resonance. The QCDSF \cite{Schierholz:LAT09} and the ETM Collaborations \cite{Feng:2009ck} reported on their projects to calculate $m_\rho$ by the above method.

Results for the $\rho$ decay constant have been obtained by several groups, see the overview in Tab.~\ref{tab:rhoDecay}. Listed there are the decay constants from the coupling to the vector current and the tensor current, $f_\rho$ and $f_\rho^T$, resp., in the $\overline{\rm MS}$ renormalization scheme at 2 GeV. Experimentally observed are $f_\rho\simeq 208\,{\rm MeV}$ from $\tau^-$ decay \cite{Becirevic:2003pn} and $f_{\rho^0}\simeq 216(5)$  from $\rho^0\to e^+e^-$. 

%
%
\begin{table}
\begin{center}
\begin{tabular}{lcc}
\hline\hline
                                           & $f_\rho\:[{\rm MeV}]$ & $f_\rho^T\:[{\rm MeV}]$ \\\hline
ETMC \cite{Jansen:2009hr}                  & 239(18)              & 159(8) \\
RBC-UKQCD \cite{Allton:2008pn}             &                      & 143(6) \\
Hashimoto et al.\ \cite{Hashimoto:2008xg}  & 210(15)              &        \\
QCDSF \cite{Gockeler:2005mh}               &                      & 168(3) \\\hline\hline
\end{tabular}
\end{center}
\caption{\label{tab:rhoDecay}The vector meson decay constants $f_\rho$ and $f_\rho^T$ (renormalized in $\overline{\rm MS}$, $\mu=2\,{\rm GeV}$).}
\end{table}



\section{Excited states}

The extraction of properties of baryon states beyond the ground state is more demanding for several reasons. First of all, the signal has to be obtained from the sub-leading exponentials in the fit to the correlator, having a much weaker statistical signal. Since in most cases a straight-forward multi-exponential fit would fail for that reason, one has to come up with a more sophisticated approach better suited to the problem at hand. The variational approach \cite{Michael:1985ne,Luscher:1990ck} seems to be most successful to solve this problem, for other approaches and a comparison see, e.g., \cite{Lin:2007iq} and references therein. In the variational approach instead of fitting a single correlator one uses a whole matrix constructed from several correlators. These correlators need to have a sufficient overlap with states one intends to extract. Possibilities to construct several such correlators are, e.g., using different operators or different smearing prescriptions for the fields. Here I will focus on the mass spectrum of excited light hadron states only. 

\begin{figure}
\begin{center}
\includegraphics[width=.55\textwidth]{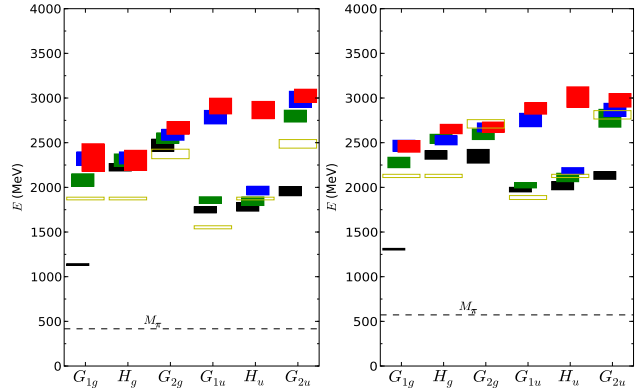}
\end{center}
\vspace*{\closercaption}
\caption{\label{fig:excitedNucleon}Energy spectrum of excited nucleon states from Bulava et al.\ \cite{Bulava:2009jb}, obtained with $N_f=2$ flavors of anisotropic Wilson fermions at light meson masses of 420 MeV \textit{(left panel)} and 580 MeV \textit{(right panel)}. (Plots courtesy of J.M.~Bulava et al.)}
\vspace*{\afterFigure}
\end{figure}

Recently, J.M.~Bulava et al.\ published results on the excited state nucleon spectrum \cite{Bulava:2009jb} obtained from $N_f=2$ dynamical anisotropic Wilson fermion simulations at two different light meson masses of roughly 420 and 580 MeV. They construct operators in the irreducible representation of the octahedral group corresponding to different spins and parities and using suitable displacements of the quark fields in the baryon operators. Figure~\ref{fig:excitedNucleon} (from \cite{Bulava:2009jb}) shows their result for the $I=1/2$ baryon spectrum at the two different light meson masses (the latter is indicated on the plots by the dashed line as well). Identifying their lowest state in the positive-parity channel as the nucleon, they find a cluster of negative-parity states at around 1.5--1.7 times the nucleon mass in accordance with the pattern of physical states observed by experiment. The higher positive-parity states lie at energies of 1.8 times the nucleon or above, leaving open the question whether or not the lowest state in this cluster will come down eventually when the lightest meson masses are lowered, so that it agrees with the Roper resonance at 1.53 times the nucleon mass.  

%
%

\begin{figure}
\begin{center}
\includegraphics[angle=90, width=.65\textwidth]{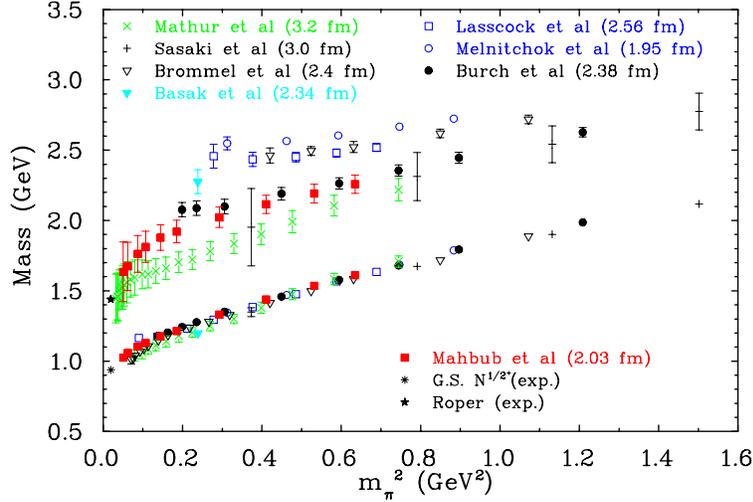}
\end{center}
\vspace*{\closercaption}
\caption{\label{fig:Roper}The nucleon ground state and Roper resonance determined from different lattice simulations. (Plot courtesy of D.~Leinweber, taken from \cite{Mahbub:2009cf}.)}
\vspace*{\afterFigure}
\end{figure}

In previous studies, the Roper resonance $N^\star(1440)$ often was difficult to isolate and/or turned out to be found at too high a mass compared to the experimentally observed mass. S.~Mahbub and collaborators in their recent work utilizing quenched FLIC fermions found a dependence of the extracted excited state mass on the smearing levels used and therefore suggested that one should extract the excitations based on the variational approach using different smearing levels in the correlator matrix \cite{Mahbub:2009nr,Mahbub:2009aa,Mahbub:2009cf}. It is their conjecture that previous work actually reported a superposition of states rather than the Roper \cite{Mahbub:2009nr}. Figure~\ref{fig:Roper} (from \cite{Mahbub:2009cf}) shows a comparison of the nucleon ground state and the Roper from various recent lattice determinations.

%
%

\begin{figure}
\begin{minipage}{.475\textwidth}
\begin{center}
\includegraphics[width=.95\textwidth]{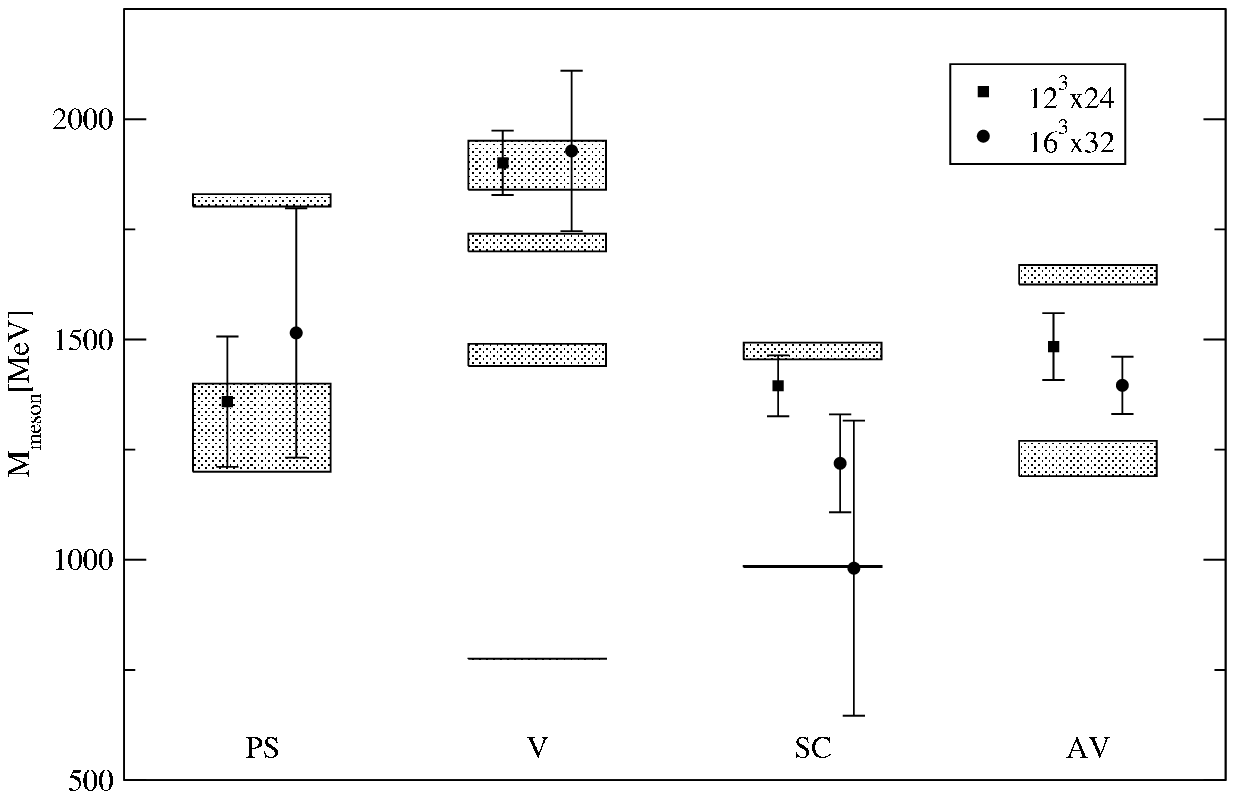}
\end{center}
\end{minipage}
\hfill
\begin{minipage}{.475\textwidth}
\begin{center}
\includegraphics[width=.85\textwidth]{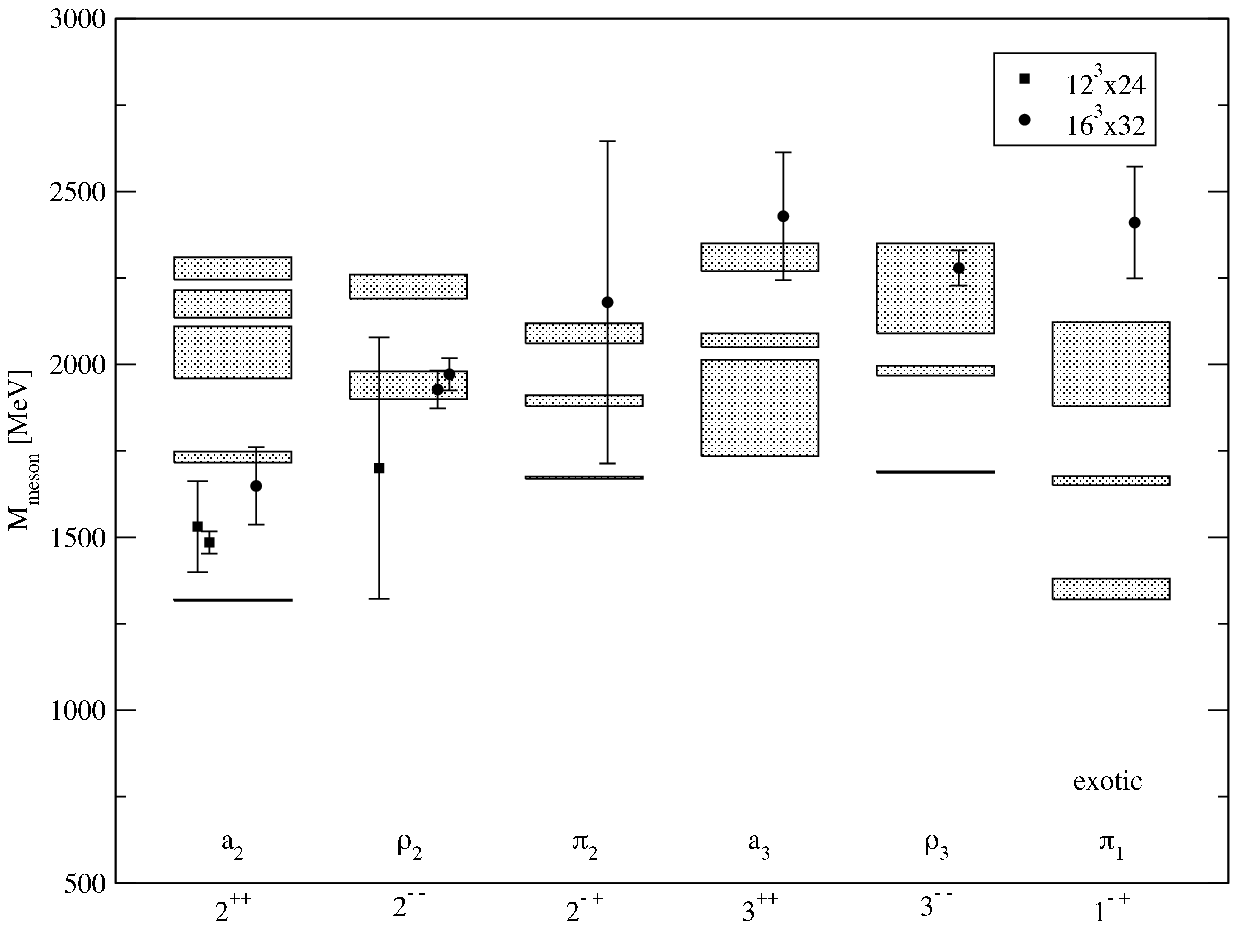}
\end{center}
\end{minipage}
%
\caption{\label{fig:excitedMeson}Energy spectrum of low (0,1) spin \textit{(left panel)} and high (2,3) spin \textit{(right panel)} meson states from Burch et al.\ \cite{Burch:2009wu}, obtained with $N_f=2$ flavors of clover-Wilson fermions at light meson masses of 500 MeV and higher. (Plots courtesy of T.~Burch et al.)}
\vspace*{\afterFigure}
\end{figure}

The meson spectrum for low and high spin excitations has been studied by T.~Burch and collaborators \cite{Burch:2009wu}. Based on $N_f=2$ dynamical clover-Wilson fermion configurations from CP-PACS with a lightest meson mass of 500 MeV, they analyzed the ground and excited states for low spin (0,1) using a variational approach with several different quark sources. For high spin (2,3), only the ground states have been extracted. See Figure~\ref{fig:excitedMeson} (from \cite{Burch:2009wu}) for a summary of their results from two different lattice spacings (0.2 fm and 0.15 fm) extrapolated to the physical point. Given the current uncertainties plus systematics from the rather high meson masses used in the extrapolation and potentially large Wilson fermion chiral symmetry breaking, at this point it might be too early to draw a definite conclusion. To overcome the limitations of chiral symmetry breaking induced by lattice artifacts, the excited meson spectrum is also examined using $N_f=2$ flavors of chirally improved (CI) fermions \cite{Gattringer:2008vj,Engel:2009cq}, where good signals have been obtained in the meson sector. The baryon spectrum from the CI fermion study currently turns out to give masses which are too high. This might be caused by the lattice volume being too small to produce reliable results for baryons.

%
%

There is an open question, whether or not the lightest scalar mesons, the $\sigma$, $\kappa$, and $a_0(980)$ are tetra quark ($\bar{q}\bar{q}qq$) states. S.~Prelovsek et al.\ are currently investigating the $I=0$, $1/2$, $3/2$, and $2$ tetra quark channels by using dynamical CI fermion configurations and also quenched overlap fermions \cite{Prelovsek:2008rf,Prelovsek:2009bk}, finding indications for a strong tetra quark component in the case of the $\sigma$ and $\kappa$ states.


\section{Concluding remarks}

This review covered recent and current efforts of the lattice QCD community to extract the masses and decay constants of the light hadrons from numerical simulations following a non-perturbative, first-principle approach to QCD. In the light pseudo-scalar sector over the recent years steady progress in simulating lighter quark masses approaching the physical point has been made. Also the understanding of the advantages and limitations of the chiral extrapolation has gained a lot of insight from this progress and I tried to highlight the current status and discussion of chiral fits for the pseudo-scalar meson sector. For the future, the inclusion of the complete NNLO terms should be further pursued, both to improve the precision of the fits and to gain more insight into the convergence of the chiral expansion. This year the first promising attempts in this direction have been presented but all require additional phenomenological input, which eventually should be avoided. First results available close to or reweighted to the physical masses offer now the possibility to test the values predicted by the extrapolations. The light hadron spectrum as summarized in Fig.~\ref{fig:hadronSpect} beautifully demonstrates the success of lattice QCD showing that the many different fermion discretizations used, combine into a consistent ``big picture'', although here some issues about the extrapolation methods used in the baryon sector need to be sorted out in the future. The study of excited states on the lattice also looks very promising. The tools for the extraction are well understood and first results with dynamical fermion simulations have been obtained.



{\small \noindent\textbf{Acknowledgments} I would like to thank the organizers of the ``Lattice 2009'' conference in Beijing for their invitation to give this plenary talk. Many colleagues helped me with this task by providing their material and preliminary results, promptly and patiently answering my questions and participating in fruitful discussions. I would like to thank %
Constantia Alexandrou, %
Yasumichi Aoki, %
Claude Bernard, %
Tom Blum, %
Peter Boyle, %
Tommy Burch, %
Norman~H.\ Christ, %
Christine Davies, %
Chris Dawson, %
Xu Feng, %
Hidenori Fukaya, %
Steven Gottlieb, %
Thomas~R.\ Hemmert, %
Gregorio Herdoiza, %
Roger Horsley, %
Chulwoo Jung, %
Takeshi Kaneko, %
Chris Kelly, %
Yoshinobu Kuramashi, %
John Laiho, %
Christian~B.\ Lang, %
Derek Leinweber, %
Laurent Lellouch, %
Huey-Wen Lin, %
Matthew Lightman, %
Vittorio Lubicz, %
Chris Maynard, %
Robert~D.\ Mawhinney, %
Craig McNeile, %
Jun-Ichi Noaki, %
Shigemi Ohta, %
Haralambos Panagopoulos, %
Sasa Prelovsek, %
Siebren Reker, %
Christopher Sachrajda, %
Gerrit Schierholz, %
Amarjit Soni, %
Cecilia Tarantino, %
Carsten Urbach, %
and my colleagues in the RBC-UKQCD Collaboration, the Fermilab Lattice Group, and at the University of Regensburg. For suggestions and a careful reading of the manuscript I am thankful to Norman H.~Christ and Christopher Sachrajda.

I gratefully acknowledge the financial support provided by Fermi National Accelerator Laboratory, operated by Fermi Research Alliance, LLC, under Contract No.~DE-AC02-07CH11359 with the United States Department of Energy. }


\bibliography{references}

\bibliographystyle{JHEP-2} 


\end{document}